# Stable Operation of AlGaN/GaN HEMTs at 400°C in air for 25 hours

Saleh Kargarrazi[1], *Member, IEEE,* Ananth Saran Yalamarthy[2], *Member, IEEE,* Peter F. Satterthwaite[3], Scott William Blankenberg[4], Caitlin Chapin[2], and Debbie G. Senesky[1], *Member, IEEE,*

***Abstract*—In this letter, we report the operation of AlGaN/GaN HEMTs with Pd gates in air over a wide temperature range from 22°C to 500°C. The variation in the threshold voltage ($V_{th}$) is less than 1% over the entire temperature range. Moreover, a safe biasing region where the transconductance peak ($g_m$) occurs over the entire temperature range was observed, enabling high-temperature analog circuit design. Furthermore, the operation of the devices over 25 hours was experimentally studied, demonstrating the stability of the DC characteristics and $V_{th}$ at 400°C. Finally, the degradation mechanisms of HEMTs at 500°C over 25 hours of operation are discussed, and are shown to be associated with the 2DEG sheet density and mobility decrease.**

## I. Introduction

Wide-bandgap semiconductors such as silicon carbide (SiC) and gallium nitride (GaN) have proven to be viable candidates for operation in extreme environments due to their superior electronic properties, namely their low intrinsic carrier concentration [1], [2]. Wide-bandgap electronics have recently found use in aviation, space exploration, automotive and deep-well drilling applications. NASA Glenn Research center (GRC) has studied SiC JFETs (junction field effect transistors) for harsh environments for approximately a decade [3] and demonstrated reliable electronics operational at 500°C for a year [4]. High-temperature devices and circuits in SiC have also been studied to a great extent in n-type metal-oxide semiconductor (NMOS) and complementary metal-oxide semiconductor (CMOS) [5]–[8], and BJT (bipolar junction transistor) architectures [9]–[18]. AlGaN/GaN high electron mobility transistor (HEMTs) are a promising candidate for implementation of integrated electronics [19]–[22], and a few studies have discussed the high-temperature capability of the HEMTs [23]–[25]. However, as compared with SiC, AlGaN/GaN HEMTs have not been sufficiently matured for implementation of ICs for extreme temperatures.

Previous work has investigated the high-temperature properties of depletion-mode AlGaN/GaN HEMTs, at temperatures of 200°C [26], 400°C [27], and 425°C [28]. Maier et al. [24] reported the failure of the GaN HEMTs with $Al_{0.24}Ga_{0.76}N$ barrier layer grown on SiC, with Ti/Al/Ni/Au Ohmic contacts and Mo/Au Schottky gate contact after 219 hours at 500°C in vacuum (benign environment). The corresponding failure was associated with the gate diode breakdown. Furthermore, the DC characteristics of AlGaN/GaN HEMTs and MIS-HEMTs have been studied up to 600°C in air for 30 minutes [29]. It is shown in [29] that the HEMTs fail prematurely at 300°C due to the high gate leakage. Although these MIS-HEMTs operate up to 600°C, they exhibit threshold voltage ($V_{th}$) instability beyond 300°C.

In this letter, we investigate high-temperature operation of depletion-mode AlGaN/GaN HEMTs on Si substrates with Pd gates. The operation of the HEMTs from 22°C up to 400°C is studied. The high temperature stability of HEMTs kept at 400°C and 500°C for 25 hours is further analyzed and discussed. It is shown that the AlGaN/GaN HEMTs exposed to 400°C environments have a stable response over 25 hours of operation with $\sim$ 1% variation of the threshold voltage $V_{th}$ after 5-hours burn-in, and can be promising candidates for future integrated circuits. Moreover, we show that the prolonged exposure to temperatures above 400°C in air can limit the operation of the AlGaN/GaN HEMTs due to the reduction of the 2DEG (2D electron gas) sheet density and mobility. Compared with [29], we report improved gate leakage and DC characteristics for the HEMTs and unlike [24] that investigated the high-temperature operation of the AlGaN/GaN HEMTs in vacuum, this letter demonstrates the high-temperature AlGaN/GaN HEMTs in the oxidizing environment of air. Reliable operation in air is important for practical applications as it avoids complex packaging and enables sensing applications where contact with the ambient is necessary [30].

## II. AlGaN/GaN HEMT Fabrication

The HEMTs were fabricated with an AlGaN/GaN-on-Si wafer (DOWA, Inc.) grown by metal-organic chemical vapor deposition (MOCVD). The cross-section is illustrated in Fig. 1. It consists of a 1.5 $\mu$m thick strain management buffer structure followed by a 1.5 $\mu$m thick GaN layer grown on top of Si (111). Formation of the 2DEG was accomplished by growing an epitaxial stack consisting of a 1-nm-thick AlN spacer, 30-nm-thick $Al_{0.25}Ga_{0.75}N$ barrier layer and 1 nm thick GaN capping layer. This wafer has a manufacturer specified 2DEG mobility of $\sim$1,400 cm$^2$/V-s and sheet density of $\sim$1x10$^{13}$ cm$^{-2}$ at room temperature. A mesa etch is used to define the HEMT channel via an inductive coupled plasma technique with $BCl_3/Cl_2$ gases. Source/drain contacts were realized using a standard evaporation and lift-off process with a Ti/Al/Pt/Au (20/100/40/80 nm) stack. To make the contacts Ohmic, rapid thermal annealing (RTA) was employed at 850°C for 35 s. A layer of Pd/Au (40/10 nm) was e-beam evaporated and patterned as the gate metal. The devices were passivated by a 20 nm thick atomic layer deposited (ALD) $Al_2O_3$ layer, deposited at 250°C [31]. The contact pads were opened up by etching the ALD $Al_2O_3$ using a 20:1 buffered oxide etch (BOE) solution for 1 min. Ti/Pt (10/100 nm) is used for the interconnect/bond pads (Fig. 1). Prior to measurement all samples were annealed at 600°C for 30 seconds in air on a hot-chuck.

[1] Affiliated with the Department of Aeronautics and Astronautics, Stanford University, Stanford, CA 94305 USA. [2] Affiliated with the Department of Mechanical Engineering, Stanford.[3] Affiliated with the Department of Electrical Engineering and Computer Science, Massachusetts Institute of Technology (MIT).[4] Affiliated with the Department of Electrical Engineering, Stanford. (Corresponding author: Saleh Kargarrazi). The authors would like to acknowledge the support from Knut and Alice Wallenberg Foundation, Stanford Nanofabrication Facility (SNF), and advice from Prof. Jim Plummer.

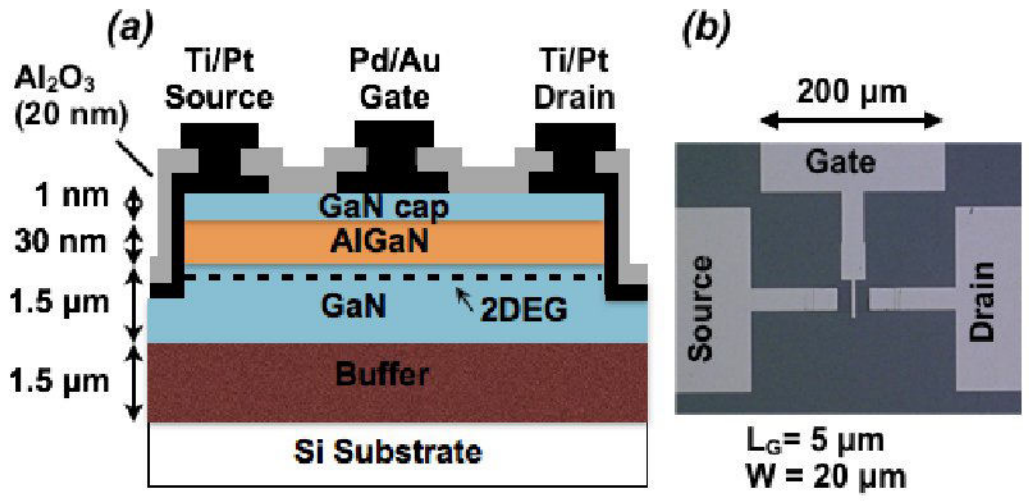

Fig. 1. (a) The cross section schematic and (b) the micro-photograph of the AlGaN/GaN HEMT ($L_{GS}$=$L_{GD}$=6 $\mu$m).

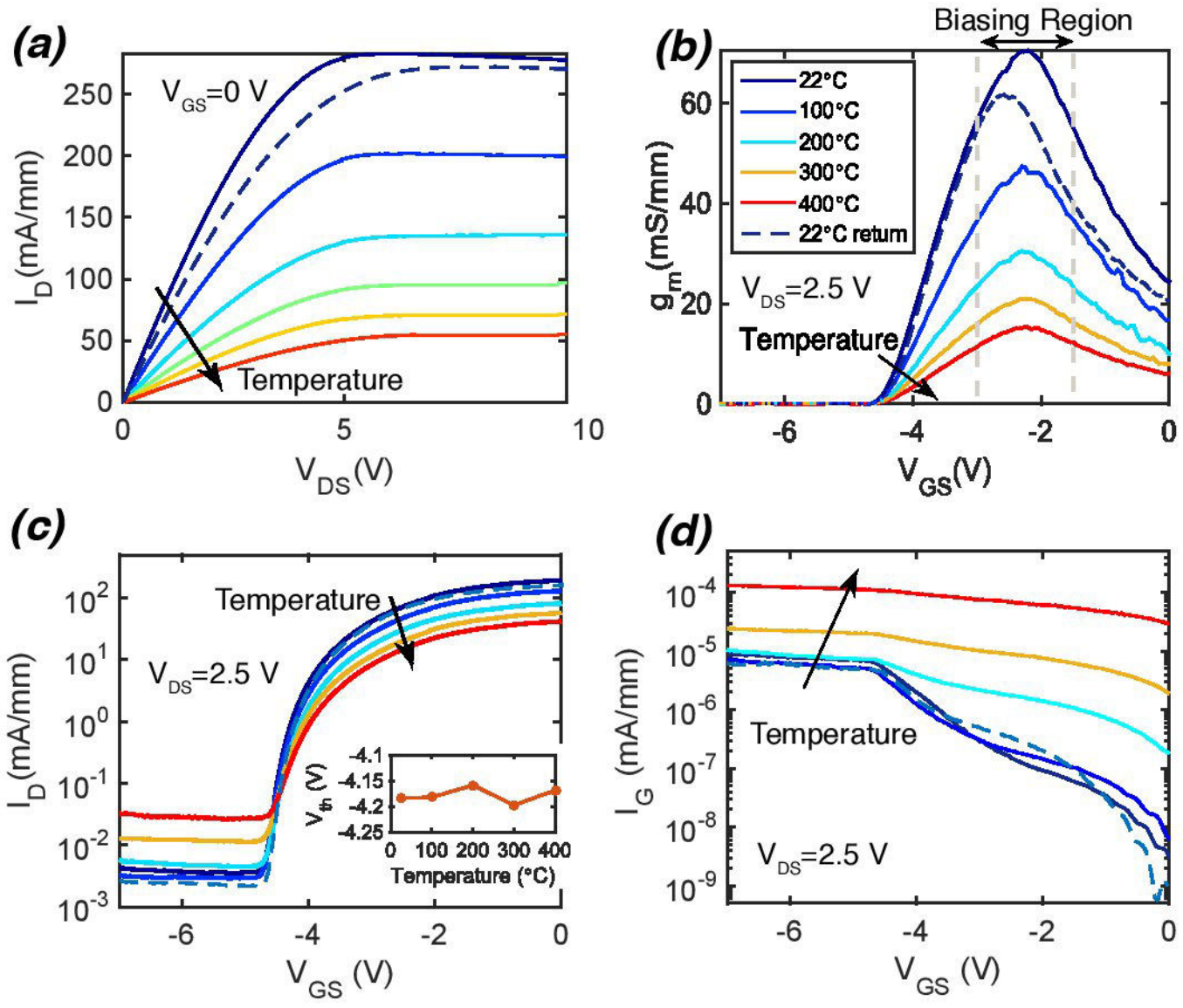

Fig. 2. (a) The output characteristics (at $V_{GS} = 0$ V). (b) The transconductance ($g_m$) at $V_{DS} = 2.5$ V. (c) The $I_D - V_{GS}$ (at $V_{DS} = 2.5$ V). The inset shows the temperature dependence of $V_{th}$. (d) The gate leakage current of the GaN HEMT over temperature from 22°C to 400°C.

## III. Experimental Results and Discussions

The GaN HEMTs were characterized in air from 22°C up to 500°C on a temperature-controlled probe station using an Agilent B1500A semiconductor parameter analyzer. The measurement at 22°C was repeated after cooling and denoted as 22°C return in Fig. 2.

### *A. DC Characteristics of the AlGaN/GaN HEMT*

The output characteristics ($I_D - V_{DS}$) and transconductance ($g_m$) of the HEMTs in the temperature range of 22°C to 400°C are shown in Fig. 2(a) and (b), respectively. The decrease in drain current ($I_D$) and $g_m$ is due to lowered mobility at high temperatures. This decrease was observed to have a $T^{-1.5}$ dependence on temperature, consistent with the decrease in 2DEG electron mobility from optical phonon scattering [32]. The peak of the transconductance ($g_{m,peak}$) shifts from -2.3 V to -2.5 V as the temperature is increased from 22°C to 400°C and decreased upon return to 22°C. This peak position defines an optimal biasing region for analog design. The proposed biasing region for a temperature-stable analog amplifier design is indicated in Fig. 2(b). The existence of a $\sim$1.5 V biasing region across the measured temperature range means that these devices are suitable for high-temperature analog design.

The HEMT characteristics are partly recovered after cooling to room temperature. However, both $I_D$ and $g_{m,peak}$ decrease slightly, and the $g_m$ shifts to more negative voltages. Fig. 2(c) illustrates $I_D - V_{GS}$, showing the decrease of the HEMT *ON/OFF* ratio from $\sim 5\times10^4$ at room temperature to $\sim 10^3$ at 400°C. The threshold voltage ($V_{th}$) of the HEMTs have been extracted using the *Extrapolation in the Linear Region* (ELR) method [33], considering long-channel MOSFET behavior. The variation of $V_{th}$ over temperature is quite small ($\sim$ 0.9%) as shown in the inset of Fig. 2(c). $V_{th}$ after return to room temperature was measured as -2.52 V. Moreover, the gate leakage current ($I_G$) of the HEMT over the temperature is observed to degrade monotonically over temperature (from $10^{-5}$ to $10^{-4}$), but recovered as the sample is cooled down to 22°C (Fig. 2(d)).

### *B. Prolonged Measurements at 400°C and 500°C (25 hours)*

Electrical characterization of AlGaN/GaN HEMTs over prolonged exposure to high temperature was performed. Devices were placed on a hot-chuck in air for 25 hours, at 400°C and then 500°C. The DC characteristics of the HEMTs were recorded in 5-minute intervals. The devices measured in this experiment had previously been tested intermittently at elevated temperatures up to 500°C. Fig. 3(a)-(c) depict the output characteristics, the $I_D - V_{GS}$, and the $V_{th}$ of the HEMTs at 400°C over the course of 25 hours. In all the aforementioned plots, little degradation is observed over time. The $I_D - V_{GS}$ characteristics stabilized after the 5 hours of operation at 400°C, as can be seen in Fig. 3(b). No variation of $V_{th}$ from hour 5 to hour 25 was observed (Fig. 3(c)). However, when the temperature was increased to 500°C, measurements shows obvious degradation of the DC characteristics over time (Fig. 3(d)-(f)). In these devices, $n_{sh}$ is given by [34]

$$n_{sh} = \frac{\sigma_{tot}}{q} - \frac{\epsilon_0 \cdot \epsilon_{AlGaN}}{q^2 \cdot d}(q\phi_b + E_f - \Delta E_C), \qquad (1)$$

where $\sigma_{tot}$ is the total polarization charge (including both spontaneous and piezoelectric components), $\phi_b$ the Schottky barrier height, $E_f$ the Fermi level with respect to the GaN conduction-band-edge energy at the GaN/AlGaN interface, $\Delta E_C$ the conduction band discontinuity at the AlGaN/GaN interface, $q$ the elementary charge, $d$ the thickness of the AlGaN barrier layer, $\epsilon_0$ the permittivity of air, and $\epsilon_{AlGaN}$ the permittivity of AlGaN.

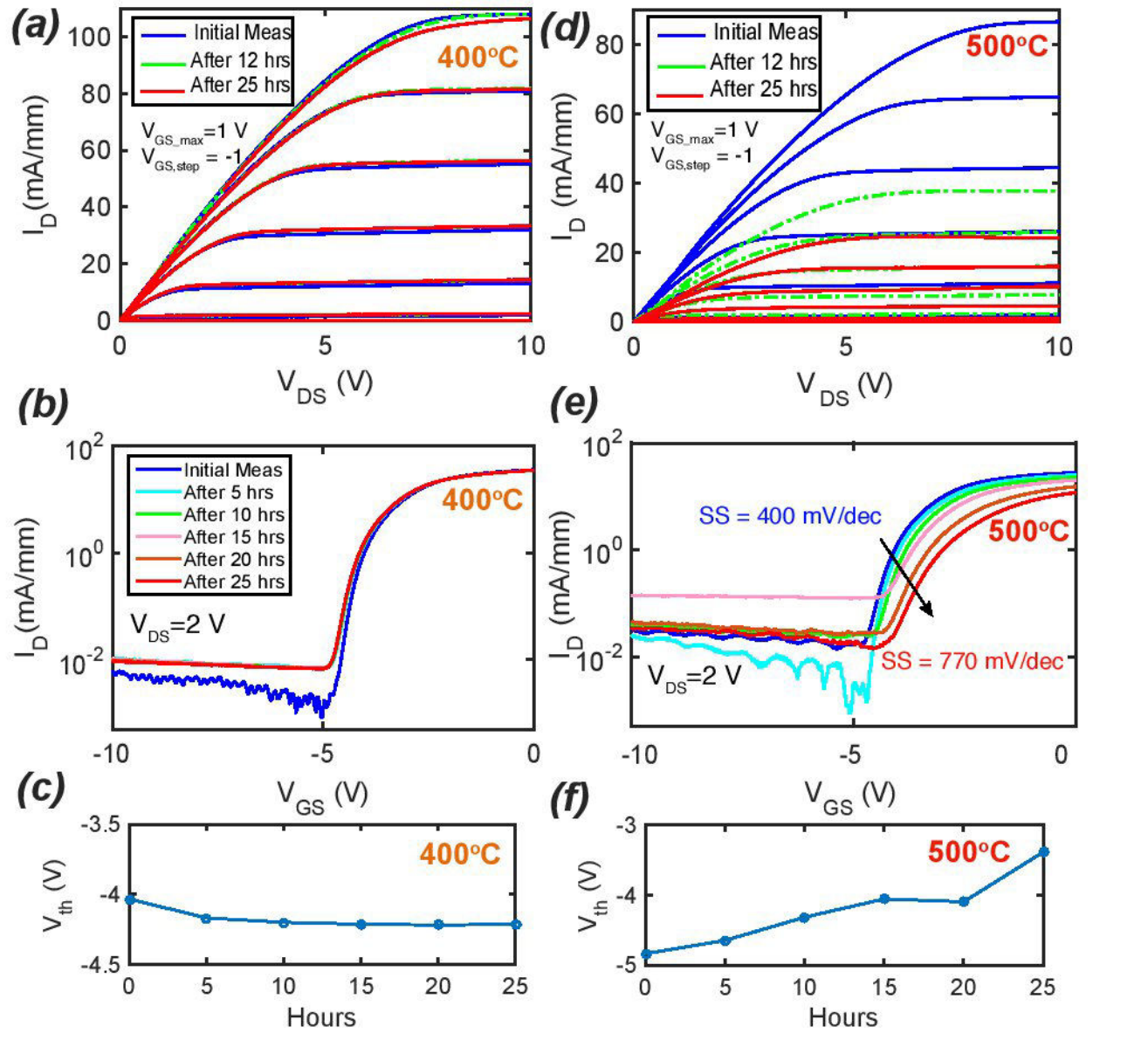

Fig. 3. (a), (d) the output characteristics, (b), (e) $I_D - V_{GS}$ (at $V_{DS} = 2$ V), and (c) ,(f) the threshold voltage of the GaN HEMT (at $V_{DS} = 2$ V), at 400°C and 500°C over the course of 25 hours of measurement (the legend for (b) and (e) are the same).

The degradation of the HEMT performance at 500°C over time is partly related to the reduction in the sheet density of the 2DEG ($n_{sh}$). It occurs due to *strain relaxation*, where a reduction in strain in the AlGaN barrier layer results in reduced piezoelectric polarization [35], [36]. The gate leakage current ($I_G$) at 500°C shows little variation ($\sim 5\times10^{-6}$) over 25 hours. This indicates that $\phi_b$ can be assumed to be constant over time. The variation of $E_f$ and $\Delta E_C$ over time is also assumed negligible compared to the variation of $\sigma_{tot}$. In contrast to the measurements at 400°C, a monotonous reduction of $|V_{th}|$ is observed over time at 500°C as shown in Fig. 3(e),(f). The $|V_{th}|$ in a depletion-mode HEMT is given by [34]:

$$|V_{th}| = \frac{q \cdot n_{sh} \cdot d}{\epsilon_0 \cdot \epsilon_{AlGaN}} \tag{2}$$

In this expression only $n_{sh}$ has a strong temperature dependency. The drastic reduction of $|V_{th}|$ at 500°C over time thus supports the role of *strain relaxation* in device degradation. If we assume that the drain current in the linear region follows a long-channel model, where $I_D \propto \mu[(V_{GS} - V_{th})(V_{DS}) - V_{DS}^2]$, and the mobility ($\mu$) does not change, we can estimate a $\sim$1.6$\times$ degradation in $I_D$ from $V_{th}$ degradation over the 25-hour duration. However, $\sim$2.3$\times$ reduction of *ON*-state $I_D$ is observed $V_{GS} = 0$ V (Fig. 3.(e)) which indicates the 2DEG mobility decreased by a factor of $\sim$1.4$\times$ over time at 500°C, besides the decrease in the 2DEG sheet density. The mobility degradation may also be associated with cracks developed at the AlGaN/GaN interface due to strain relaxation [36]. Moreover, the subthreshold slope shows a degradation over time from 400 $mV/$decade to 770 $mV/$decade (Fig. 3.(e)), which is presumed to be associated with the interface traps, similar to Si MOSFETs [37]. Furthermore, the association of subthreshold slope change with interface traps in AlGaN/GaN HEMTs is introduced and studied in [38].

## IV. Conclusions

High-temperature operation of AlGaN/GaN HEMTs with a Pd gate was demonstrated for 25 hours of operation, from 22°C to 500°C. $V_{th}$ variation is less than 1% over the entire temperature range. The peak of the transconductance, ($g_{m,peak}$) shifts from $V_{GS}$= -2.8 V to -2.3 V, suggesting a safe biasing operation region for analog electronics over a wide range of temperature 22°C to 400°C. Moreover, 25-hour measurements of the HEMTs show stable DC characteristics at 400°C, with a very slight degradation of $I_D$, and unchanged $V_{th}$. The HEMTs at 500°C exhibit a degradation behaviour over 25 hours which is attributed to the reduction of the 2DEG sheet density as well as mobility due to strain relaxation.